\documentstyle[aas2pp4]{article}
\lefthead{Jessner et al.}
\righthead{Charge densities above pulsar polar caps}
\begin{document}
\title{Charge densities above pulsar polar caps}
\author{Axel Jessner\altaffilmark{1}, Harald
Lesch\altaffilmark{2} and Thomas Kunzl
\altaffilmark{2,3}}
\altaffiltext{1}{Max-Planck-Institut f\"ur Radioastronomie, Auf dem
H\"ugel 69, D-53121 Bonn, Germany}
\altaffiltext{2}{Universit\"ats-Sternwarte M\"unchen, Scheinerstr.1,
D-81679 M\"unchen, Germany}
\altaffiltext{3} {Max-Planck-Institut f\"ur extraterrestrische
Physik,Giessenbachstr., D-85740 Garching, Germany}

\authoraddr{A. Jessner}
\authoremail{jessner@mpifr-bonn.mpg.de}
\date{}
\setcounter{footnote}{0}
\begin{abstract}
The ability of the neutron star surface to supply all or only part of the
charges filling the pulsar magnetosphere is crucial for the physics prevailing 
within it, with direct consequences for the possible formation of pair creation
regions. We evaluate the conditions for $e^-$ emission from pulsar surfaces
for  a simple Goldreich-Julian geometry taking both thermal and field emission
processes into account. Using recently published estimates for the equation of
state at the neutron star's surface, we show, that for a large range of $T_{\rm
surf}, B$ and $P$, the liberated charges will fully screen the accelerating
B-parallel electric field  ${\rm E_\|}$. For surface temperatures $T_{\rm
surf}<2\cdot 10^5 $K a balance between field emission of electrons and
shielding of the field will occur. Even in the overidealised case of  $T_{\rm
surf}=0$ one can expect a prodigious supply of electrons which will weaken the
accelerating ${\rm E_\| }$. We calculated the motion of electrons along selected
polar field lines numerically for the low temperature, field emission scenario
yielding their Lorentz factors as well as the produced radiation
densities. Inverse Compton  and later curvature losses are seen to balance the
acceleration by the residual electric fields.  We found that the conditions for
magnetic pair production are not met anywhere along the field lines up to a
height of 1500 pulsar radii. We did not {\it a priori} assume an "inner gap",
and our calculations did not indicate the formation of one under realistic
physical conditions without the introduction of further assumptions.

\end{abstract}

\section{Introduction}

A proper assessment of the  conditions very close to the surface of a neutron
star is a vital ingredient for any model that attempts to explain the enigmatic
origin of the observed electromagnetic radiation from pulsars.  

Numerous studies of the polar cap acceleration have been made within a
particular framework: A steady acceleration of electrons within a space charge
limited particle flow originating at the surface of the neutron star will
produce particles emitting magnetically convertible curvature $\gamma-$rays
(e.g. Arons \& Scharlemann 1979). 
These high energy photons are converted into $e^+e^-$-pairs with high enough
efficiency (Daugherty \& Harding 1982) so that in the end, they and their
daughter products will exceed the initial (Goldreich-Julian density) by a factor
of $10^4$ (e.g. Zhang \& Harding 2000). An important {\it a priori } assumption
is inherent in all these calculations: Above a critical height $h_{\rm gap}$ a
region of high conductivity will quench any electric field ${\rm E_\|}$ parallel
to the guiding B-field lines, so that the full potential across the polar cap
will be active between the outer surface of the neutron star and the inner
surface of the high conductivity region. 

Our approach differs in that respect from other workers, we do not make that 
initial assumption, but try to calculate the amount of charges produced by the
surface, how they affect and are affected by  ${\rm E_\|}$, their energy losses
and the possibility of pair production. Should one find suitable conditions for
such a pair production region to exist, the methods quoted above will
be the prime choice for its description.
 
The conclusion that electron-positron pairs with high densities exist close to
the neutron star has recently been questioned by several authors. Kunzl et
al. (1998) considered this widespread assumption of dense pair 
plasmas for the problem of radio wave propagation within the neutron star
magnetosphere.
On hand of the most recent emission height estimates (50 - 100 $R_{\rm NS}$) for
the radio emission (Kijak \& Gil (1997); Kramer et al. (1997, 1998)) they found
that the existence of low frequency radio emission for example from the Crab
pulsar is incompatible with the scenario of extended dense pair
production. These findings were fully
confirmed by the theoretical investigation of Melrose \& Gedalin (1999), who
treated the plasma emission mechanisms in detail and came to the conclusion
that only low energy particles with Goldreich-Julian density could exist
in the radio emission regions in order to alow for escaping radio waves
(see also Melrose (2000) and references therein). Furthermore
recent simulations of the pair cascade by Arendt \& Eilek (2000) for strong
and weak magnetic fields shed some light on the difficulties for an
efficient production of secondary pairs by high energetic primary particles via
$\gamma$-ray curvature photons. Although their findings need some more
investigation they could show that the field strength at the surface has the
most important effect on the cascade. For strong fields 
$B\sim 10^9\,{\rm T}$ they find almost no secondary photons and the cascade
terminates in one generation. The weak fields ($B\sim 10^7\,{\rm T}$) had either
no cascades or quite weak ones. Only for moderate fields ($B\sim 10^8\,{\rm T}$)
they observed some secondary particle generations. Arendt \& Eilek concluded
that it is not yet clear whether pulsars can really produce a large number of
pairs at the surface gaps.

This tempted us to question the canonical {\it a priori} assumption 
of the existence of an inner pair creating gap and investigate its possibility
given a reasonable estimate of the surface characteristics and the fields
involved. 

In the literature there are three fundamentally different ideas for pair
production in the inner magnetosphere: A "starved" magnetosphere where the
neutron star surface provides only an insufficient supply of charged particles
(e.g. Ruderman and Sutherland (1975)), a magnetosphere where the
Goldreich-Julian density is reached at the surface but due to curvature effects
of the field lines deviations occur close to the surface (e.g. Arons \&
Scharlemann (1979), Barnard \& Arons (1982)) and Shibata's model which predicts
electric fields caused by {\em overdense} regions, that means in the opposite
sense than in the Arons model (Shibata, 1997). 

The latter two ideas suffer from  serious weaknesses since it turned out that
Arons type models do not work for slow pulsars (Beskin (2000) gives an upper
limit of 0.1 to 0.3 s for the period). The Shibata model in turn neglects the
back reaction of an overdense region which reduces the particle flow from
inside. Therefore we will restrict our discussion to models of
Ruderman-Sutherland type and fix the parameter range in which this kind of
models is reasonable. 

For the scope of this article we consider a strong magnetic dipole anchored in
a heavy ($1.4 M_{\rm sun}$) sphere with a radius $r= 10\,{\rm km}$ that rotates
twice per second ($\Omega=2 \pi \cdot 2 s^{-1}$).
 Let the B-field have a strength of $|B_0|=10^{8}~{\rm T}$ at the surface and
for simplicity`s sake let the magnetic and the rotational axes be
aligned\footnote{We use spherical coordinates $(r,\theta,\phi)$ centered in the  
neutron star with $\theta$ being the directional angle referred to the pole
which coincides with the rotation axis.}.
\begin{equation}
  {\rm B}(r,\theta) = B_0\cdot \left( {r_{\rm NS} \over r}\right) ^{-3}
  \cdot (\cos\theta ,{1\over 2}\sin\theta , 0)
\end{equation}
Such objects have been discussed as idealised cases of radio pulsars by
Goldreich \& Julian (1969)
and many authors since then. Of course one has to
realise that the above mentioned idealisations limit the applicability of
such simple models for the interpretation of observational facts
from the growing number of pulsars that is being discovered.
We will however try to extend the scope of these models by constraining
the number of charged particles available and
analysing the fate of individual charged particles
(limiting ourselves to electrons and positrons) in the polar regions
of such heavy rotating dipoles under the constraints given by some of the
more recent observations.  Although a pulsar magnetosphere is characterized
by extreme physical conditions in terms of strong electric and magnetic fields
the fundamental processes that can supply charges
are reasonably well known
from terrestial experiments and technical applications and can be
extrapolated to the surface conditions of a neutron star. Such an extrapolation
provides constraints on the initial charge density which is vital for any 
model of cascading particle creation and the radiation processes.

\section{Accelerating fields and charge densities }
Muslimov \& Tsygan (1990, 1992) showed how the electric and magnetic
fields of a rotating neutron star are distorted by relativistic effects.
A comprehensive treatment of the rotating dipole in the frame of general 
relativity was given by Muslimov \& Harding (1997). For our purposes of 
{\it demonstrating} the major physical properties near the polar cap it 
suffices however to stick to a classical description (including special 
relativity of course).

Our rotating magnetic dipole acts as an unipolar generator and induces an
 electric field as the consequence of its surface charge distribution
\begin{eqnarray}
  {\rm E}(r,\theta) &=& -B_0 \Omega r_{\rm NS} 
  \left({r\over r_{\rm NS}}\right)^{-4} \cdot\nonumber\\
  &\cdot& (3\cos^2\theta-1, 2\cos\theta \sin\theta,0).
\end{eqnarray}

The component parallel to the magnetic field lines
${\rm E_\|={{E\cdot B}\over |B|}}$
is  given by
\begin{equation}
{\rm E_\|}(r,\theta)= -4 B_0 \Omega r_{\rm NS} \left({r \over r_{\rm NS}}
\right)^{-4}{\cos^3\theta \over \sqrt{3\cos^2\theta+1}}
\end{equation}
and reaches $2.5\cdot 10^{13}~{\rm V/m}$ at the surface near the
magnetic poles (Goldreich \& Julian 1969; Holloway 1973). In vacuum
all appropriately charged particles will be quickly accelerated to
very high energies, but because of the strong magnetic field they
 are constrained to move along the magnetic field lines given by the parametric
equation
\begin{equation}
r(\theta,\theta_0)=r_{\rm NS}\cdot {\sin^2\theta \over \sin^2\theta_0}.
\end{equation}
Here $\theta_0$ is meant to be the colateral angle of the field line at
the surface of the sphere.

Goldreich \& Julian (1969) already calculated the electron density which 
shields the electric field and yields the only stable static solution for a 
pulsar magnetosphere (excluding centrifugal and gravitational forces).

\begin{equation}
 n_{\rm GJ}(r,\theta) = {{\varepsilon_0 \Omega B_0} \over e}
\left({r \over r_{\rm NS}}\right)^{-3}\cdot (3\cos^2\theta - 1)
\end{equation}
with $e$ being the electron charge.

In our example this amounts to
\mbox{$n_{\rm GJ0}={{2\varepsilon_0 \Omega B_0} \over e}=$} 
$1.4\cdot 10^{17}{\rm m^{-3}}$ at the poles on the surface.
Deutsch (1955) was the first who showed
that the high induced fields make it unlikely that stellar rotators with
strong magnetic fields will rotate in a (charge-free) vacuum.
Several well studied processes enable i.e.~electrons to escape from a
metallic surface.
If the neutron star surface can thus liberate sufficient charges to keep
the Goldreich-Julian current flowing, the emission will be space-charge 
limited like an electronic valve (diode). Electrons will be accelerated 
away\footnote{The observed high brightness temperatures of
the pulsar radio emission suggest the operation of beaming mechanisms
for the enhancement of the radio intensity. We'll therefore focus
our attention on outflowing particles where intensity enhancement by
forward beaming would be natural.} from the polar
regions if $\vec{\Omega}\cdot \vec{B} >0 $\footnote{In some models in the literature we need
{\em anti-alignment} of $\vec{\Omega}$ and $\vec{B}$. In that case the
Goldreich-Julian charge density above the polar cap is positive. We do not
discuss this case here but since the binding energies of ions are assumed to be
comparable to those of electrons the qualitative argumentation remains
unchanged.}. 

Usov \& Melrose (1995)
have reminded us of the fact that the relevant fundamental surface emission
processes are field emission and thermal emission (perhaps enhanced by the
Schottky effect (Schottky 1914)).
Thermal surface emission of electrons was discovered by T.A. Edison (1884) and
has been extensively studied since the beginning of our century (Dushman 1930).
Field emission from metal surfaces is also known for a long time  (Fowler \&
Northeim 1928) and used in many technical applications.
The effectiveness of both emission processes is controlled by $E_{\rm W}$, the
work function or binding energy, the energy required
to liberate an electron from the crystal lattice at the surface boundary.
Values for $E_{\rm W}$ and the Fermi-energy $E_{\rm F}$ for the pulsar surface can
only be extrapolated from terrestial values as the high densities
of $\rho_{\rm Fe} \approx 10^6 {\rm kg~m^{-3}}$ and magnetic fields 
$B \approx 10^{8}$T
cannot be reached in laboratory experiments. We will therefore first
try to establish a {\it reasonable range of values} for $E_{\rm W}$ in order
to investigate the effectiveness of surface emission.

\subsection{Constraints for the range of the work function}

Using a Thomas-Fermi-Dirac-Weizs\"acker approximation
Abrahams \& Shapiro (1992) calculated the  cold $(T=0)$
surface density $\rho_{\rm Fe}$
to be about $2.9\cdot 10^6\, {\rm kg m^{-3}}$. 
That result enables us to give an improved estimate
of the Fermi energy at the surface:

\begin{equation}
E_{\rm F}(\rho_{\rm Fe}) ={{2\cdot \pi^4\hbar^4c^2} \over {e^2B^2m_e}}
\cdot\left({{\rho_{\rm Fe}\cdot(28-2)}\over 56\cdot m_p}\right)^2
\end{equation}

which amounts to $E_{\rm F} =4.17\cdot 10^2\, {\rm eV}$ 
(as usual $\hbar =h/2\pi$). In our subsequent 
calculations we take that as the most likely value of $E_{\rm F}$. In the case of an
unknown work function $E_{\rm W}$, one usually approximates $E_{\rm W}=E_{\rm F}$.
So far calculations of the actual electron work function have  not been made
and for want of a better alternative we will use the Fermi energy as the best
available measure of the work function.   Abrahams \& Shapiro (1992) give the
ionisation energy for iron as 120 {\rm eV} and
we will consider the range from ionisation energy to cyclotron energy
(0.12 - 12 {\rm keV}) to be also a reasonable constraint for $E_{\rm W}$.

\subsection{Field (cold cathode) emission}
 By calculating the
 transmission coefficient of electron wave functions through a potential
wall \newline $U(x)=e^{-1} E_{\rm W}-{\rm E_0}\cdot x$ for $x>0$ and $U(x)=0$ for 
$x<0$ (${\rm E_0}$ being the electric field) Fowler \& Nordheim (1928) 
provided a description of electron currents from metal surfaces in good 
agreement with the experimental results and useful in many technical 
applications since then. Using their expression of the cold cathode 
current density and assuming a relativistic flow of electrons we find 
for the particle density


\begin{equation} 
n_{\rm field}(E_0) =
\frac{e^2}{2\pi h c }
\frac{\sqrt{E_{\rm W} \over E_{\rm F}}\cdot  E_0^2} {(E_{\rm W} + E_{\rm F})} 
 {\rm e}^{-\frac{8\pi\cdot\sqrt{2\cdot m_e}}{3 h e}
\cdot \frac{E_{\rm W}^{3/2}}{E_0}}
\end{equation}

In Fig. 1 we plot the ratio of the field emission charge density to the
Goldreich-Julian density for a standard pulsar. The authors like to 
emphasize the point that this density is independent of the somewhat
 uncertain surface temperature and depends only on $E_0$.

\begin{figure}[htbp]
\label{fig1}
\input{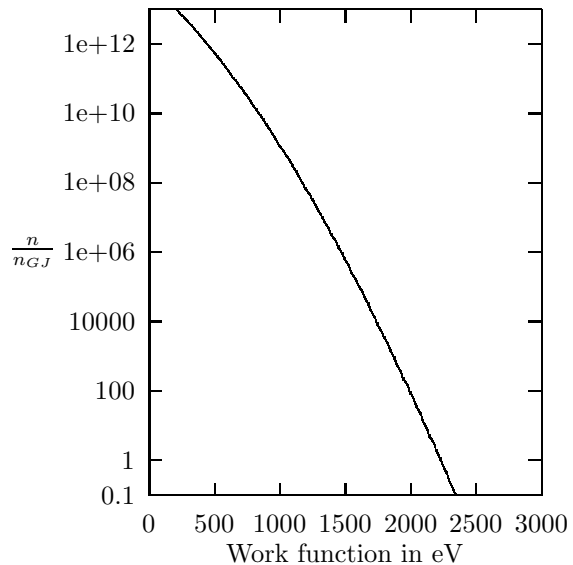}
\caption{Cold cathode emission $n/n_{\rm GJ}$ versus work function $E_{\rm W}$
for a standard pulsar (see text).}
\end{figure}

Field emission can supply the Goldreich-Julian density for a cold
neutron star surface for any $E_{\rm W} < 2 {\rm keV}$! Fig. 2 however 
shows, that for the extreme case of a 100 times weaker field
(i.e. a slowly rotating weak pulsar) the maximum binding energy drops down
as far as 50 {\rm eV}.

\begin{figure}[htb]
\label{fig2}
\input{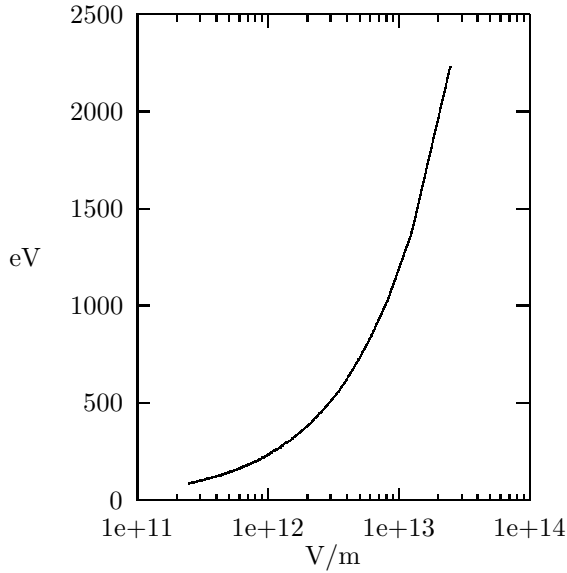}
\caption{The critical work function for the Goldreich-Julian density versus the
surface field ${\rm E_0}$.}
\end{figure}

Fig. 3 shows how for an assumed work function of i.e. 500 {\rm eV},
the cathode efficiency increases dramatically with the electric field.
We may therefore expect field emission to play a significant role in
the supply of charges for the average pulsar magnetosphere.
\begin{figure}[htb]
\label{fig3}
\noindent
\setlength{\unitlength}{0.240900pt}
\ifx\plotpoint\undefined\newsavebox{\plotpoint}\fi
\begin{picture}(990,720)(0,0)
\font\gnuplot=cmr10 at 10pt
\gnuplot
\sbox{\plotpoint}{\rule[-0.200pt]{0.400pt}{0.400pt}}%
\put(220.0,113.0){\rule[-0.200pt]{4.818pt}{0.400pt}}
\put(198,113){\makebox(0,0)[r]{0.1}}
\put(906.0,113.0){\rule[-0.200pt]{4.818pt}{0.400pt}}
\put(220.0,155.0){\rule[-0.200pt]{4.818pt}{0.400pt}}
\put(198,155){\makebox(0,0)[r]{1}}
\put(906.0,155.0){\rule[-0.200pt]{4.818pt}{0.400pt}}
\put(220.0,238.0){\rule[-0.200pt]{4.818pt}{0.400pt}}
\put(198,238){\makebox(0,0)[r]{100}}
\put(906.0,238.0){\rule[-0.200pt]{4.818pt}{0.400pt}}
\put(220.0,322.0){\rule[-0.200pt]{4.818pt}{0.400pt}}
\put(198,322){\makebox(0,0)[r]{10000}}
\put(906.0,322.0){\rule[-0.200pt]{4.818pt}{0.400pt}}
\put(220.0,405.0){\rule[-0.200pt]{4.818pt}{0.400pt}}
\put(198,405){\makebox(0,0)[r]{1e+06}}
\put(906.0,405.0){\rule[-0.200pt]{4.818pt}{0.400pt}}
\put(220.0,488.0){\rule[-0.200pt]{4.818pt}{0.400pt}}
\put(198,488){\makebox(0,0)[r]{1e+08}}
\put(906.0,488.0){\rule[-0.200pt]{4.818pt}{0.400pt}}
\put(220.0,572.0){\rule[-0.200pt]{4.818pt}{0.400pt}}
\put(198,572){\makebox(0,0)[r]{1e+10}}
\put(906.0,572.0){\rule[-0.200pt]{4.818pt}{0.400pt}}
\put(220.0,655.0){\rule[-0.200pt]{4.818pt}{0.400pt}}
\put(198,655){\makebox(0,0)[r]{1e+12}}
\put(906.0,655.0){\rule[-0.200pt]{4.818pt}{0.400pt}}
\put(220.0,113.0){\rule[-0.200pt]{0.400pt}{4.818pt}}
\put(220,68){\makebox(0,0){1e+12}}
\put(220.0,677.0){\rule[-0.200pt]{0.400pt}{4.818pt}}
\put(345.0,113.0){\rule[-0.200pt]{0.400pt}{4.818pt}}
\put(345,68){\makebox(0,0){2e+12}}
\put(345.0,677.0){\rule[-0.200pt]{0.400pt}{4.818pt}}
\put(510.0,113.0){\rule[-0.200pt]{0.400pt}{4.818pt}}
\put(510,68){\makebox(0,0){5e+12}}
\put(510.0,677.0){\rule[-0.200pt]{0.400pt}{4.818pt}}
\put(636.0,113.0){\rule[-0.200pt]{0.400pt}{4.818pt}}
\put(636,68){\makebox(0,0){1e+13}}
\put(636.0,677.0){\rule[-0.200pt]{0.400pt}{4.818pt}}
\put(761.0,113.0){\rule[-0.200pt]{0.400pt}{4.818pt}}
\put(761,68){\makebox(0,0){2e+13}}
\put(761.0,677.0){\rule[-0.200pt]{0.400pt}{4.818pt}}
\put(926.0,113.0){\rule[-0.200pt]{0.400pt}{4.818pt}}
\put(926,68){\makebox(0,0){5e+13}}
\put(926.0,677.0){\rule[-0.200pt]{0.400pt}{4.818pt}}
\put(220.0,113.0){\rule[-0.200pt]{170.075pt}{0.400pt}}
\put(926.0,113.0){\rule[-0.200pt]{0.400pt}{140.686pt}}
\put(220.0,697.0){\rule[-0.200pt]{170.075pt}{0.400pt}}
\put(45,405){\makebox(0,0){$n\over n_{GJ}$}}
\put(573,23){\makebox(0,0){$V \over m$}}
\put(220.0,113.0){\rule[-0.200pt]{0.400pt}{140.686pt}}
\put(800,661){\usebox{\plotpoint}}
\multiput(796.56,659.92)(-0.913,-0.499){133}{\rule{0.829pt}{0.120pt}}
\multiput(798.28,660.17)(-122.279,-68.000){2}{\rule{0.415pt}{0.400pt}}
\multiput(673.43,591.92)(-0.649,-0.499){111}{\rule{0.619pt}{0.120pt}}
\multiput(674.71,592.17)(-72.715,-57.000){2}{\rule{0.310pt}{0.400pt}}
\multiput(599.92,534.92)(-0.499,-0.498){101}{\rule{0.500pt}{0.120pt}}
\multiput(600.96,535.17)(-50.962,-52.000){2}{\rule{0.250pt}{0.400pt}}
\multiput(548.92,481.47)(-0.498,-0.638){77}{\rule{0.120pt}{0.610pt}}
\multiput(549.17,482.73)(-40.000,-49.734){2}{\rule{0.400pt}{0.305pt}}
\multiput(508.92,430.17)(-0.497,-0.729){63}{\rule{0.120pt}{0.682pt}}
\multiput(509.17,431.58)(-33.000,-46.585){2}{\rule{0.400pt}{0.341pt}}
\multiput(475.92,381.74)(-0.497,-0.860){53}{\rule{0.120pt}{0.786pt}}
\multiput(476.17,383.37)(-28.000,-46.369){2}{\rule{0.400pt}{0.393pt}}
\multiput(447.92,333.26)(-0.496,-1.006){45}{\rule{0.120pt}{0.900pt}}
\multiput(448.17,335.13)(-24.000,-46.132){2}{\rule{0.400pt}{0.450pt}}
\multiput(423.92,284.95)(-0.496,-1.104){39}{\rule{0.119pt}{0.976pt}}
\multiput(424.17,286.97)(-21.000,-43.974){2}{\rule{0.400pt}{0.488pt}}
\multiput(402.92,238.56)(-0.495,-1.222){35}{\rule{0.119pt}{1.068pt}}
\multiput(403.17,240.78)(-19.000,-43.782){2}{\rule{0.400pt}{0.534pt}}
\multiput(383.92,192.09)(-0.495,-1.370){31}{\rule{0.119pt}{1.182pt}}
\multiput(384.17,194.55)(-17.000,-43.546){2}{\rule{0.400pt}{0.591pt}}
\multiput(366.92,145.73)(-0.493,-1.488){23}{\rule{0.119pt}{1.269pt}}
\multiput(367.17,148.37)(-13.000,-35.366){2}{\rule{0.400pt}{0.635pt}}
\end{picture}
\caption{The cold cathode efficiency  $n/n_{\rm GJ}$ at $E_{\rm W}=417 {\rm eV}$ as a 
function of the surface field ${\rm E_0}$.}
\end{figure}
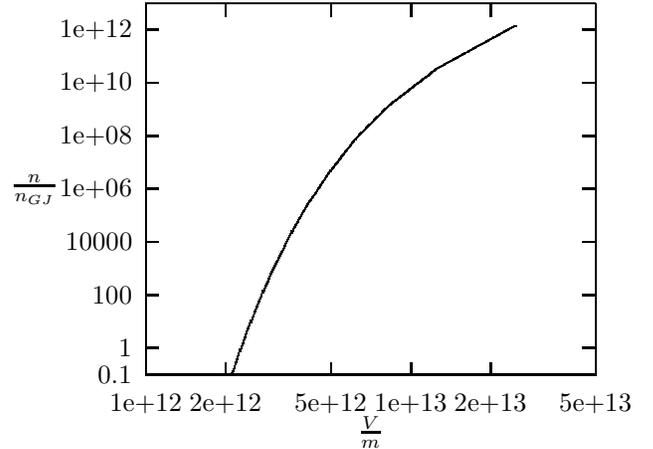

\subsection{The thermal electron emission}

In contrast to the quantum mechanical effect of field emission, the thermal
surface emission of electrons from a conductor is a purely classical
effect: For a given particle distribution function, be it Maxwellian or 
Fermi--Dirac, there will be a fraction of particles with a kinetic energy that
exceeds that of the potential barrier characterised by the binding energy.
That number of electrons liberated from a hot metallic surface is
given by  Dushman's equation (Dushman 1930). In the presence of
strong external electric fields, the number of electrons is increased
over that given by Dushman's equation because of the Schottky effect
(Schottky 1914) which lowers the effective potential barrier
for the particles.
The charge density is thus described by the combined Dushman-Schottky-Equation

\begin{equation}
n_{\rm DS}(T,E_0)= {A_0\over e\cdot c}\cdot T^2 \cdot
{\rm e}^{-{E_{\rm W}\over k T}+{e\over k T}\cdot\sqrt{{e\cdot E_0}\over 
{4\pi \epsilon_0}}}
\end{equation}
(k being the Boltzmann constant) with 
$A_0={{4\pi\cdot m_e\cdot e\cdot k^2}\over h^3}$ $= 120.2\, 
{\rm A cm^{-2}}$  which depends only on $E_{\rm W}$.
In Fig. 4 we show the contributions to the current density according from
field emission and from thermal emission with and without the Schottky
effect for a range of temperatures likely for a standard pulsar with
an assumed $E_{\rm W}=417$\,{\rm eV}.
\begin{figure}[htb]
\label{fig4}
\input{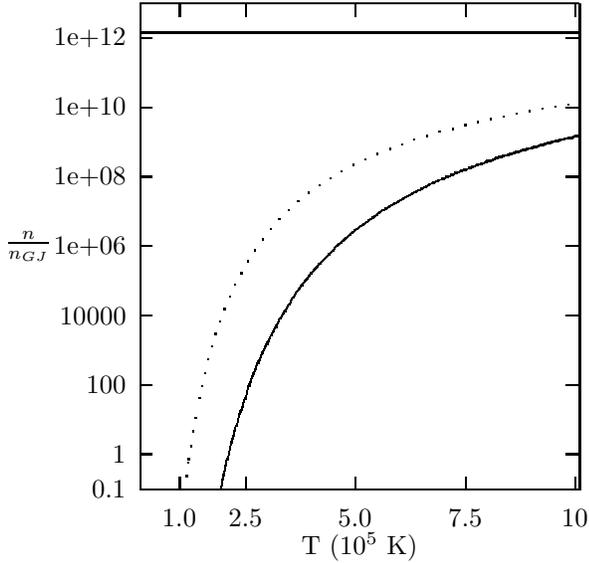}
\caption{The surface efficiency  $n/n_{\rm GJ}$  at $E_{\rm W}=417 {\rm eV}$ for
a standard pulsar as a function of the surface temperature. Thin solid line:
thermal emission, dots: thermal emission with Schottky effect, Thick solid: cold
cathode emission.} 
\end{figure}

As has been
pointed out before, for 
\mbox{$E_0=2.5\cdot10^{13}\,{\rm {V/m}}$}
field emission alone can supply all that
is required, but for temperatures above $10^5\,$K, thermal emission, which is
only weakly dependent on the effective field becomes even more efficient.
From $2\cdot 10^5\,$K onwards, thermal electrons, emitted from the surface even
without a large electric field can supply the GJ-density. Fig. 5 brings
everything into perspective by outlining the extremes of the parameter space.

\begin{figure}[htb]
\label{fig5}
\input{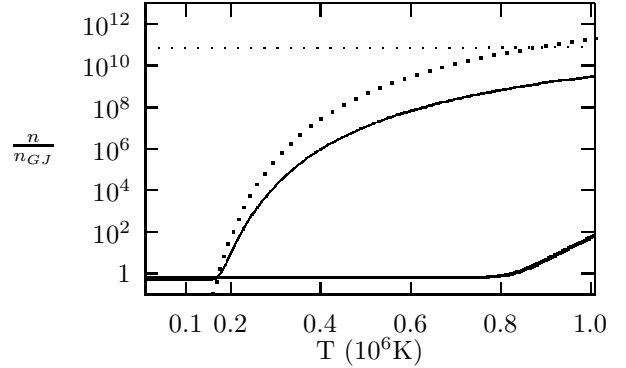}
\caption{The surface efficiency  $n/n_{\rm GJ}$  for different  $E_{\rm W}$ and as a function
of  the surface temperature.  long dash:  total efficiency for $E_{\rm W}=417 {\rm eV}$,
solid:  total efficiency for $E_{\rm W}=417 {\rm eV}$  and ${\rm 0.1 E_0}$,
short dash:  total efficiency for $E_{\rm W}=2065{\rm eV}$,
dots:   total efficiency for $E_{\rm W}=417 {\rm eV}$ and a weak
 field pulsar with ${\rm 0.01 E_0}$ and ${\rm 0.01 n_{\rm GJ}}$ .}

\end{figure}

We show the effective ratio of total charge emission for work functions of
$E_{\rm W}=2065{\rm eV}$ and $E_{\rm W}=417\,${\rm eV} as well as for a field of 
${\rm E_0} \over 100$. Only for extremely high $E_{\rm W}\gg 1200\,${\rm eV} or very 
low fields $E<10^{12}{\rm V/m}$ and low surface temperatures $T< 2\cdot 10^5$K 
the electron supply from the pulsar surface will be insufficient to fill the 
magnetosphere with charges to shield the induced electric field. It is however 
doubtful that objects with such low temperatures can be found as internal 
stresses caused by the rotational braking of the neutron star set a lower 
limit of a few $10^5$K for the surface temperature (Tsuruta 1998 and references
therein). But normally the potential supply of charges exceeds the requirements
of the G-J-density by a wide margin. Thus any accelerating electric field at 
the surface of the neutron star will be quenched within an extended corotating
charged lower region of the magnetosphere. Therefore we do not expect any
significant acceleration in the inner magnetosphere which, of course, implies
the absence of highly relativistic particles and no pair production at all under
these circumstances.

\section{Neutron stars dominated by field emission}

We have shown in the last section, that high binding energies or
low surface temperatures are the only conditions compatible with
significant surface electric field strengths that can accelerate particles.
In these cases, an equilibrium field of ${\rm E_{\rm eq} = E_0}\cdot(1-{n_e \over 
n_{\rm GJ}})$ will remain when $n_e$ electrons flow off the surface. Both the 
electric field $E_{\|}$ and the shielding density are proportional to the 
product ${\rm B}\cdot \Omega$. We introduce a scale factor 
$\xi={\rm \frac {{\rm B}\cdot\Omega}{({\rm B}\cdot\Omega)_0}}$  to account for
the possible empirical variation of ${{\rm B}\cdot\Omega}$ and by numerically 
solving the balance equation

\begin{eqnarray}
  \xi n &=& n_{\rm field}(\xi{\rm E}(1-\frac {n} {\xi n_{\rm GJ}}),E_{\rm W})+\nonumber\\
  &+& n_{\rm DS}(\xi{\rm E}(1-\frac {n} {\xi n_{\rm GJ}}),E_{\rm W},T)
\end{eqnarray}

for $0.02 < \xi < 5$  we obtain the equilibrium densities for any given 
temperature T and work function $E_{\rm W}$ covering a wide range of neutron stars
with different $B$ and $\Omega$. Fig. 6 shows the variation of the residual 
electric field $ \xi {\rm E}(1-\frac {n} {\xi n_{\rm GJ}})$ as a function of $\xi$ 
for $E_{\rm W}=417 {\rm eV}$ and $E_{\rm W}=1200 {\rm eV}$. A temperature of 
$1.85\cdot 10^5$K, just below the onset of thermal emission for $E_{\rm W}=417 
{\rm eV}$,  was chosen for the calculation. The residual field grows with 
$\xi$ until it saturates at $2.1\cdot 10^{12} {\rm  V/m}$ for $\xi=1$.  
A higher $E_{\rm W}$ causes earlier saturation at a higher level. Between 300 eV and 
2000 eV the residual field saturation level $E_{\rm lim}$ depends nearly 
linearly on $E_{\rm W}$ and can be approximated as 
$E_{\rm lim} \approx 1.3\cdot 10^{10} \cdot (E_{\rm W} -250 {\rm eV}){\rm V/m}$.
For our standard $E_{\rm W}$, the density $n$ deviates by only 10\% from the 
Goldreich-Julian value, but higher $E_{\rm W}$ show lower initial densities with 
the expected increase in charge density with increasing electric fields.

\begin{figure}[htb]
\label{fig6}
\input{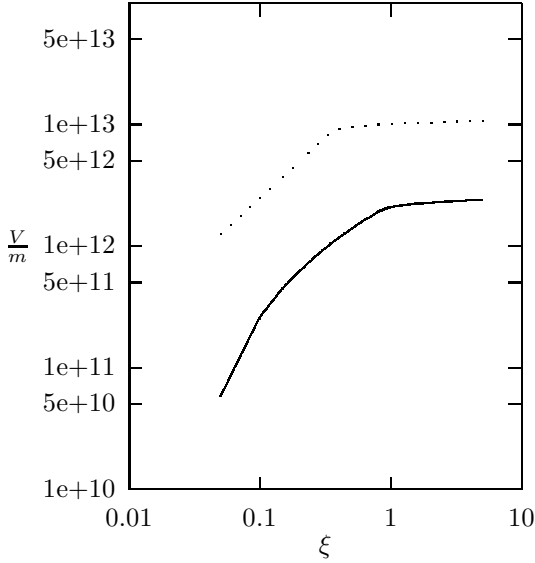}
\caption{The residual electric surface field  
\mbox{$ \xi {\rm E}(1-\frac {n} {\xi n_{\rm GJ}})$} 
at $T = 1.85\cdot 10^5$K 
as a function of $\xi$
for $E_{\rm W}=417 {\rm eV}$ (solid line) and $E_{\rm W}=2065 {\rm eV}$ (dots).}
\end{figure}

\subsection{The equilibrium energy of electrons in the magnetosphere}

Any residual field will quickly accelerate the electrons to relativistic
energies, balanced by the braking due to various energy loss processes. 
Because of the high magnetic field strength, the motion of charges will be 
along the field lines. With the help of the parametrisation (4) we can 
describe a particle trajectory using only its co-lateral angle at the pulsar 
surface (starting angle $\theta_s$) and the co-lateral angle $\theta$ on the 
trajectory. At each point of such a  trajectory, the energy gain of a charge 
by the parallel component of the residual electric field

\begin{equation}
{d\gamma \over dt}\left|_{\rm E}\right. = -\left({e \over m_e c} \right) 
(1-{n_e \over n_{\rm GJ}})
{\rm E_\|}(\theta,\theta_0)
\end{equation}

will be balanced by a number of energy loss mechanisms. Curvature radiation is
an obvious loss mechanism for relativistic charged particles forced to follow a
curved trajectory. The energy loss of a relativistic particle due to curvature
radiation is given by (e.g. Zheleznyakov 1996, p.231)
\begin{equation}
  {d\gamma \over dt}\left|_{\rm CR}\right. = \frac{2}{3}\gamma^4\frac{e^2}
  {4\pi\varepsilon_0 m_e c}\frac{1}{R_{\rm C}(\theta,\theta_0)^2}
\end{equation}
Here we use the curvature radius of a dipolar field line
in its parameterised form (Lesch et al. 1998)

\begin{equation} R_c(\theta,\theta_0)={
{r_{\rm NS}\over 3}
{\sin\theta \over \sin^2\theta_0} {{(1+3\cos^2\theta)^{3/2}} \over 
{1+\cos^2\theta} }}
\end{equation}
The curvature radii in the
polar regions of a neutron star with a dipolar magnetic field are of the 
order of $10^6$m. To balance the
huge acceleration of an unshielded field at the surface
${d\gamma \over dt}\left|_{\rm E}\right. \approx 1.5\cdot 10^{16}{\rm s}^{-1}$
by curvature radiation with
${d\gamma \over dt}\left|_{\rm CR}\right. \approx 8.5\cdot 10^{-19} \gamma^4 
{\rm s}^{-1} $ one needs Lorentz factors of $\gamma \approx 3.6\cdot 10^8$. 
Particles with smaller Lorentz factors experience  only insignificant losses 
due to the $\gamma^{-4}$ dependence of the curvature losses.

Inverse Compton scattering of relativistic particles on the thermal photons 
emitted by the hot neutron star surface (Sturner 1995)  limits  
the electrical acceleration of particles near the pulsar surface (Supper \&
Tr{\"u}mper, 2000). If one also includes the interaction of the charged
particles with their own production of scattered photons one would expect an even more efficient braking to take place. Due to the curvature of their trajectories and
we expect a fairly strong interaction of the particles with their own previously
upscattered radiation within a few additional meters of travel, so that the
effective energy density of the radiation field will increase quickly in the
particle frame. 

For an assumed complete dissipation of the available energy of the 
residual electric field near the surface we can estimate a effective 
temperature of
\begin{eqnarray}
T &=&\left({\varepsilon_0 {\rm E_\|}^2
(\xi-{n\over n_{\rm GJ}})^2}\over a \right)^{1\over 4} = \nonumber\\
&=& 7.3\cdot 10^7{\rm K}(\xi-{n\over n_{\rm GJ}})^{1\over 2} x^{-2}
\end{eqnarray}
as a natural limit for the produced radiation.
Here we use the blackbody constant 
$a= 7.566\cdot 10^{-16}{\rm Jm}^{-3}{\rm K}^{-4}$ and 
$x = {r\over r_{\rm NS}}$ as the normalised 
distance from the surface. It is however unrealistic, to expect such efficient 
conversion of the available energy into radiation. Following Sturner (1995) we 
expect dissipation scales for the dominant processes of the order of a few m 
in the vicinity of the neutron star surface. Again a simple estimate from a 
balance of the energy gain due to the electrical field on a dissipation scale 
$\lambda$ will give us an improved temperature limit
\begin{eqnarray}
T &=& \left(e{\rm E_\| }(1-{n\over n_{\rm GJ}}) n\lambda\right)^{1\over 4} =
\nonumber\\
&=& 4\cdot 10^6{\rm K} x^{-1}
\left((1-{n\over n_{\rm GJ}}) n\lambda\right)^{1\over 4}
\end{eqnarray}

which turns out to be compatible with the observed x-ray temperatures of 
pulsars (Tsuruta 1998).

A more reliable description of the conditions near the pulsar surface can be 
attained by computing the Lorentz factor of electrons with
\begin{equation}
{d\gamma \over d\theta} = {d\gamma \over d\theta}\left|_{\rm E}\right. + 
{d\gamma \over d\theta}\left|_{\rm CR}\right. + {d\gamma(T_{\rm eff}) \over 
d\theta}\left|_{\rm invCompt}\right.
\end{equation}
together with the development of the  energy density of the radiation field
\begin{equation}
{dU \over d\theta} = - 4U {r_{\rm NS}^2\over r(\theta,\theta_0)^3} - n m_e c^2 
{d\gamma(T_{\rm eff}) \over d\theta}\left|_{\rm invCompt}\right.
\end{equation}

along a given field line parameterised by the colateral angles $\theta$ and 
$\theta_0$ from the surface out to $10^3 r_{\rm NS}$.
The energy losses by the major processes outlined by Sturner (1995)
as there are: curvature radiation
$({d\gamma \over d\theta}\left|_{\rm CR}\right.)$, the Klein-Nishina 
correction as well as magnetic, resonant and non-resonant inverse Compton 
scattering, were included in the term 
${d\gamma(T_{\rm eff}) \over d\theta}\left|_{\rm invCompt}\right.$. 
The self-enhanced 
scattering from the radiation produced in lower levels of the magnetosphere 
was approximated by a local calculation of the effective temperature of the
radiation bath $T_{\rm eff}= ({3U\over a})^{1/4}$. We did not calculate the exact spectrum
of the radiation bath for reasons of efficiency. But we expect our crude approximation to be valid
due to the curved trajectories of the electrons and the straight, tangential propagation of the emitted photons. After a short propagation, the scattering will not be back on, but at an slowly increasing angle.
We repeated the calculations also without this enhancement effect, only to find, that the initial energy plateau did not extend quite as high up into the magnetosphere and the maximum energy achievable
by the particles was a little higher.  
 Because of the widely varying 
scales of the problem, the integration was performed for twelve segments of 
the trajectory using an Bulirsch-Stoer algorithm with decreasing resolution 
from one segment to the next (Press et al. 1992). The final values of each
segment served as initial conditions for the subsequent ones. 
The results of each segment were carefully checked for their resolution 
independence, too coarse resolutions showing up as either oscillatory 
behaviour or non-convergence.
Because of the enormous acceleration of a particle starting with
$\gamma=1$ at the surface, the first (surface-) segment had to be as
short as 10 cm subdivided into 200 steps. Most interesting was the
limiting case of $T=2\cdot 10^5$ K and $n = 0.9 n_{\rm GJ}$. Fig. 7 provides
 a plot of the achieved Lorentz factors and effective temperatures
 as a function of distance from the
surface. After travelling about 5 cm, the acceleration was balanced by
non-resonant inverse Compton losses at $\gamma=2.2\cdot 10^5$.
After that initial peaking of the Lorentz factors, the particle
energies decreased slowly down to $\gamma=1.1\cdot 10^5$ at $h=20 {\rm m}$
 acccompanied by a slow rise of the effective temperature. 
That decrease is a consequence of self-enhancement and does not happen without
it. Over the next $630 {\rm m}$ the effective temperature rises quickly to $10^6
{\rm K}$ and the Lorentzfactor reaches $1.3\cdot 10^6$. Further, slower progress
leads to a complete balance and saturation at $h=1.3\cdot 10^5 {\rm m}$ with
$T=2.9\cdot 10^6 {\rm K}$ and $\gamma=1.1\cdot 10^8$. At 
$5.5\cdot 10^4{\rm m}$ curvature radiation takes over as the dominant energy 
loss mechanism. The Lorentz factors now decrease slowly down to 
$\gamma=3.9\cdot 10^7$ at a distance of $1.5\cdot 10^7 {\rm m}$.

\begin{figure}[htb]
\label{fig7}
\input{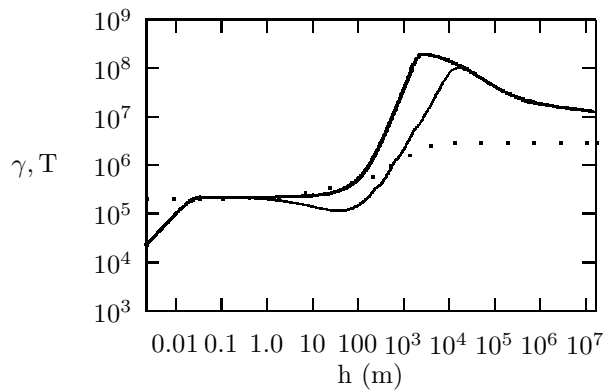}
\caption{The  Lorentz factors (thin solid line: with self-enhanced inverse Compton scattering, 
thick solid: without enhanced scattering) and effective temperatures (dashed
line) for enhanced scattering as a function of distance from the surface for $T=2\cdot 10^6$K and  
$n =0.9\cdot n_{\rm GJ}$ .}
\end{figure}

A number of different calculations with varying initial parameters
$n < 0.1 \cdot n_{\rm GJ}$
and T were made, but showed  no qualitative difference in the behaviour of 
their solutions. The case described can hence be taken as a reasonable 
representative of a "starved" system, having a low surface temperature 
and producing a strong source of X- and $\gamma $- rays with 
$T_{\rm eff} \approx 3\cdot 10^6 {\rm K}$.

\subsection{Pair production}
Our calculations differ from those of Sturner (1995) in as far
as we did not {\it a priori} assume the existence of an inner polar gap.
It is often unquestionly assumed that after about $100$m a particle creation
avalanche creates a dense $n \gg n_{\rm GJ}$ charge neutral plasma that shorts the
available electric field. For an estimate of the location of such an avalanche
region we calculated the magnetic pair creation probability for curvature 
photons of a characteristic energy
\begin{equation}
E_{\rm curv} = {{3ch\gamma^3} \over {4\pi R_c}}
\end{equation}
and Compton boosted thermal radiation having an energy
of
\begin{equation}
E_{\rm comp} = \gamma^2 k T_{\rm eff}
\end{equation}
The hard photons were assumed to be emitted tangentially and travelling outward
 from the field lines over a distance $\lambda$, their angle with the magnetic 
field lines $\Theta(\lambda,\theta,\theta_0)$ and the corresponding component
of $B$ perpendicular to the path of the photon
\begin{equation}
B_{\bot}(\lambda,\theta,\theta_0) = B_0 \cdot x^{-3} \sin\Theta(\lambda,\theta,
\theta_0)
\end{equation}
were explicitely calculated at each step along the trajectory of the
accelerated charge. We used the local strength parameter
\begin{equation}
\chi(\gamma,\lambda,\theta,\theta_0) = {\gamma\over 2} 
{B_\bot(\lambda,\theta,\theta_0) \over B_{\rm crit}}
\end{equation}
and Erber's (1966) Besselfunction approximation\footnote{Although that
approximation involves more effort for its evaluation  
as compared to the commonly used exponential form, it accurate over a greater 
range of $\chi$ and, having no discontinuities, more useful for numerical 
work.}
to calculate the local attenuation factor
\begin{equation}
\alpha_{\gamma B} = {\alpha\over \lambda_e}\cdot {0.16 \over \chi}
{\rm K_{1/3}}\left({2\over 3\chi}\right)^2
\end{equation}
with $\alpha = {1\over 137}$, the electron Compton wavelength $\lambda_e = 
{\hbar\over m_e c}$ and the critical magnetic field $B_{\rm crit} = 
{m_e^2 c^2 \over e \hbar} = 4.414\cdot 10^9 {\rm T}$.
$K_{1/3}$ stands for the modified Besselfunction of the third kind and order 
$1/3$. Multiplying the attenuation factor $\alpha_{\gamma B}$ with the local 
steplength  of the integration provides the local pair creation probability 
$p_{\gamma B}(\lambda,\theta,\theta_0)$ which in turn multiplied
with the number of emitted photons, ie.
\begin{equation}
n_\gamma = {2e^2 \over 9 c h \varepsilon_0 R_c}\gamma = 2.1\cdot 10^{-8} 
\gamma {\rm m^{-1}}
\end{equation}
 for curvature radiation, gives the local number of pairs created per 
accelerated charge\footnote{Even under favourable conditions there will be only
a small number of of initial pairs produced. This is a consequence of the fact
that $p_{\gamma  B} \leq 1$ and that, at least for curvature radiation, the
number of productive photons per unit length is very small. Because of the lower
secondary and tertiary Lorentz factors, the conditions for efficient pair
production in the form of daughter products tend to be even worse.}. We tested 
the algorithm for the physically unrealistic case of missing inverse Compton
braking and did indeed find a pair production zone at a height of 200 m above
the pulsar surface.

However, of all our simulations that included non-resonant inverse Compton 
scattering, none showed $p_{\gamma B} > 0$ anywhere near the pulsar surface or 
within the simulated range of distances, neither for curvature photons nor for 
maximally boosted inverse Compton photons. This is due to the fact that the 
calculated Lorentz factors do not reach their highest values near the pulsar 
surface but further out, where the magnetic field has already fallen 
drastically due to its $x^{-3}$ dependence.

\section{Discussion and conclusions}

We have seen that the surface temperature and the work function
play a crucial role in determining the initial energy and density of the lower
pulsar magnetosphere. Two scenarios are possible: Copious thermal emission
of electrons leading to a "saturated" charged magnetosphere with $ n = n_{\rm GJ} $ and 
${\rm E_\|} = 0$. In that case we expect the inevitable potential drop
and energy dissipation further away from the pulsar surface. Alternatively a "starved" magnetosphere, where  
$ n < n_{\rm GJ} $ is determined by the residual equilibrium field
$(1- {n\over n_{\rm GJ}}){\rm E_0}$. Here our simulations differ from those made previously
by others as we include the shielding of most of the $E_\|$ by the field
emission produced particles together with the inverse Compton loss processes.
By taking both physical effects into account, we find that conditions for
pair creation cannot be met near the surface of ordinary rotating neutron stars that
form most of the observed pulsars.

In the first case, we do not expect any Lorentz factors $\gamma > 1$.
We have, however only described an unperturbed equilibrium atmosphere.
The extensively studied fact that pulsar radio emission varies on timescales
down to microseconds (Lange et al. (1998)) can
be seen as an indication that such an
atmosphere has inherent instabilities. Theoretical examples of instabilities
in saturated magnetospheres were first given by Mestel et. al. (1985).
Because of the strength of the compensated electric field,
even smallest perturbations in the surface densities will be accompanied by
(mildly) relativistic particles. Kunzl et. al. (1999) have described a
scenario in which such perturbations lead to the radio emission via
a drift induced free electron maser process.
We expect that most neutron stars to fall into that category, having
surface temperatures higher that $2\cdot 10^5{\rm K}$ combined with an
$E_W \approx 400 {\rm eV}$.
If, as proposed by Pavlov \& Zavlin (1998) neutron stars are
surrounded by a gaseous atmosphere, we may
expect easy liberation of charges in the form of electrons and ions, and a
saturated magnetosphere is then unavoidable.

In the second category we would find cold neutron stars with a relativistic
X-ray emitting lower magnetosphere, having effective temperatures of a few 
$10^6$K. The achieved Lorentz factors are of the order of several $10^5$
in the lower magnetosphere reaching a peak of $\gamma = 10^8$ at
about $10 r_{\rm NS}$ and decrease monotonically
to a few $10^7$ at $1500 r_{\rm NS}$. A significant flow of X- and $\gamma-$rays
can be expected in these circumstances, but the authors like to emphasise
that radio emission models generally require much lower
Lorentz factors ($\approx 10 - 1000$). The discussed loss processes are
evidently incapable of braking
the charges hard enough in order to achieve this, making difficult to explain
any significant radio emission in these circumstances.

But the Lorentz factors in the region of strong magnetic fields are
nevertheless too low to expect significant pair production and the subsequent
formation of an inner gap in these cases.

But pulsars {\it are} radio and x-ray sources at the same time,
this fact can only be reconciled with the arguments above by fully 
accepting the observational evidence that the magnetosphere is far from a homogeneous or 
steady state. A global investigation of instabilities in the current system 
of neutron star, magnetosphere and its outer boundary
ought to be undertaken. Ideally, it would show that spatial or temporal
variations in the magnetosphere would create the observed x-rays  and the
radio emission in different parts of the perturbation. Our calculations are
meant to cover the possible extremes of the surface emission and the resultant
charge densities and Lorentzfactors. Our terrestial experience show us, that
surface electron emission is affected by random variations of emission
efficiency (formation of emitting islands) that are mainly responsible for the 
low-frequency part of a diode noise spectrum. It is not unthinkable that the
thermal emission efficiency fluctuates over the polar cap. In that case we can
expect local regions of the lower magnetosphere also to fluctuate individually
between the two extreme cases described.

A few self critical remarks ought to be made: This study presents a drastically
simplified scenario. We have omitted the effects of the pulsar's inclination
of the magnetic axis and possible surface inhomogeneities.
Hence a greater spatial variation between the two extremes of starved and
saturated charge liberation would be expected for real pulsars. We omitted a
treatment that includes the effects of general relativity as i.e. was done by
Muslimov and Harding (1997). If these effects are taken into account, fields and
consequently Goldreich-Julian densities will be higher. 

But due to the steep (exponential) characteristic of the surface emission
processes, one expects a shift of the critical temperature by perhaps $10^5$K.
We also neglected resonances of the pair production process described by
Daugherty \& Harding, (1996) which lead to an enhancement of the pair production
mechanism. The Lorentz factors compatible with inverse Compton braking however
are by a factor of $10^3$ lower than those constrained only by curvature radiation when pair production
would be possible near the polar cap, and complete description of the problem
would of course include both general relativistic effects and these
resonances. A more detailed investigation of the physics of the upper surface
layers would provide us with a more precise determination of $E_W$ without the
approximation of $E_W = E_F$. Furthermore, the simple approximation for $T_{\rm
eff}$ does not incorporate the detailed spectral characteristics of the produced
radiation, but the produced hard radiation will interact strongly with the
accelerated electrons, we may  even have underestimated the scattering losses.

We are lead to conclude, that the fact that we observe pulsars with
X-ray temperatures of several $10^6$K excludes the formation of an inner gap,
and the fact that radio emission is also observed from such sources
provides another argument against the high $n\gg n_{\rm GJ}$ densities that are
commonly expected from an inner avalanche region (Kunzl et al. 1998). We are
also led to conclude, that observed pulsars are dominated by free charge
emission with $n=n_{\rm GJ}$, thus it might be advantageous to study the
possible mechanisms for  radio- and X-ray emission in that context.

\bibliographystyle{unsrt}

\end{document}